# Semantic and Semiotic Interplays in Text-to-Audio AI: Exploring Cognitive Dynamics and Musical Interactions

**Guilherme Coelho**[1]

[1]**Technische Universität Berlin**

**Semantic and Semiotic Interplays in Text-to-Audio AI: Exploring Cognitive Dynamics and Musical**

**Abstract:**

This paper investigates the emerging text-to-audio paradigm in artificial intelligence (AI), examining its transformative implications for musical creation, interpretation, and cognition. I explore the complex semantic and semiotic interplays that occur when descriptive natural language prompts are translated into nuanced sound objects across the text-to-audio modality. Drawing from structuralist and post-structuralist perspectives, as well as cognitive theories of schema dynamics and metacognition, the paper explores how these AI systems reconfigure musical signification processes and navigate established cognitive frameworks.

The research analyzes some of the cognitive dynamics at play in AI-mediated musicking, including processes of schema assimilation and accommodation, metacognitive reflection, and constructive perception. The paper argues that text-to-audio AI models function as quasi-objects of musical signification, simultaneously stabilizing and destabilizing conventional forms while fostering new modes of listening and aesthetic reflexivity.

Using Udio as a primary case study, this study explores how these models navigate the liminal spaces between linguistic prompts and sonic outputs. This process not only generates novel musical expressions but also prompts listeners to engage in forms of critical and "structurally-aware listening.", encouraging a deeper understanding of music's structures, semiotic nuances, and the socio-cultural contexts that shape our musical cognition.

The paper concludes by reflecting on the potential of text-to-audio AI models to serve as epistemic tools and quasi-objects, facilitating a significant shift in musical interactions and inviting users to develop a more nuanced comprehension of the cognitive and cultural foundations of music.

**1.Introduction**

The integration of artificial intelligence with music creation has entered a transformative phase with the advent of text-to-audio models such as Riffusion (Forsgren and Martiros, 2022), MusicLM (Agostinelli et al., 2023), Stable Audio (Evans et al., 2024), Suno, and Udio. These technologies represent a paradigm shift in musical creation, introducing a novel modality that bridges linguistic structures and sound generation. This paper investigates some of the complex semantic and semiotic interplays inherent in this text-to-audio paradigm, examining its implications for musical composition, interpretation, and cognition.

Operating at the intersection of natural language processing, machine learning, and sound synthesis, text-to-audio creates a unique assemblage (Deleuze & Guattari, 1987) that challenges traditional notions of musical creativity. By transmuting descriptive language prompts into complex sound objects, these models initiate a process of semiotic transduction that echoes Hayles' (1999) concept of "intermediation" – the intricate interplay between diverse sign systems and media forms. The text-to-audio modality serves as an epistemic instrument, allowing exploration of the relationships between language, musical cognition, and cultural narratives. Users engaging with these systems actively participate in the construction and deconstruction (Derrida, 1967) of musical meaning through semantic and semiotic interactions.

The text-to-audio process can be conceptualized as an advanced form of sampling methodology, distilling linguistic signs into audial forms. It is a modality that integrates composition, mixing, performance, and both semantic and semiotic interpretation, and a shift that signifies a departure from the tactile interaction of traditional music-making towards an engagement with the cognitive and interpretative linguistic faculties of both the user and the machine. While evoking traditional sampling methods in its recontextualization of audio fragments, it extends beyond by tapping into a rhizomatic corpus of musical potentialities through latent space and training data. This approach aligns with contemporary practices such as Spawning (Herndon and Dryhurst, 2023) and older practices such as Plunderphonics (Oswald, 1985), while distinctively mining a vast expanse of musical data directed by textual nuance. Beyond mere sound production, these tools probe the semiotics of musical experience, investigating the construction and comprehension of meaning and associations within the auditory realm.

As these tools begin to permeate various platforms, exemplified by initiatives like Youtube DreamTrack[1] and viral hits such as "BBL Drizzy,"[2] (a track created using Udio that was initially misattributed as human-composed), [3] their increasing accessibility to a diverse user base precipitates a critical reexamination and change of musicking practices. This phenomenon engenders a multifaceted discourse encompassing complex considerations of musical interactions and the evolving role of the musician in the 21st century.

In engaging with these models, we assume the role of semiotic cartographers in a domain where, as Baudrillard (1981) suggests, "the map precedes the territory." We navigate a terrain shaped by structuralist thought, where human predispositions and AI's algorithmic interpretations intertwine in a complex cognitive interplay. This dynamic, illustrative of the quandaries inherent in structuralist perspectives, leads us through particular awarenesses of the musical experience.

The research explores the cognitive dynamics at play in AI-mediated musicking, including processes of schema assimilation and accommodation (Piaget, 1957), metacognitive reflection, and constructive perception - providing insights into the cognitive mechanisms underlying our engagement with these new forms of musical creation and perception.

The paper posits that text-to-audio AI models function as quasi-objects (Latour, 1993) and epistemic tools, simultaneously stabilizing and destabilizing conventional forms while fostering distinct modes of musicking (Small, 1998) and aesthetic reflexivity. Throughout the research, the author draws connections between interactions with this modality and its operational frameworks to various theories from philosophy, semiotics, and cognitive science. This interdisciplinary approach aims to contribute to the ongoing dialogue at the intersection of musicology, cognitive science, and artificial intelligence, offering new perspectives on musical expression and cognition in the age of AI-mediated creativity.

## 2. Text-to-Audio: Intersemiotic Processes and Operational Framework

### 2.1 Udio and the Text-to-Audio Modality: Linguistic-Driven Generation and Transformation of Sound

The proliferation of Large Language Models (LLMs) has accelerated the integration of text-based interactions across various domains, with music creation now joining this trend. Text-to-audio paradigm represents a fundamental departure from traditional approaches to music creation and sound manipulation. This modality uses natural language processing and machine learning to transmute textual descriptions into complex sound objects. Unlike conventional music production tools that often rely on tactile interactions, text-to-audio AI engages the cognitive and interpretative linguistic faculties of both the user and the machine, opening new avenues for exploring semantic and semiotic signification in music.

Text-to-audio systems, exemplified by models such as Udio, epitomize the convergence of linguistic interpretation and audio synthesis. Udio demonstrates a sophisticated capacity to transform descriptive text prompts into cohesive sound objects, employing a nuanced integration of semantic analysis and audio generation techniques.[45] The resulting sonic artifacts exhibit a gestalt fluency with user-specified thematic elements, faithfully embodying the composite structure of the textual input.[67] This process represents a form of intersemiotic translation, bridging the gap between verbal description and auditory realization. It underscores the potential of these systems to function as mediators in the complex interplay between language, musical cognition, and sound synthesis.

In this context, Udio functions as a intersemiotic translator, navigating what Lotman (1990) conceptualizes as the semiosphere. Lotman argues, "The unit of semiosis, the smallest functioning mechanism, is not the separate language but the whole semiotic space of the culture in question. This is the space we term the semiosphere. The semiosphere is the result and the condition for the development of culture" (Lotman, 1990). Operating within this framework, Udio negotiates between linguistic and musical sign systems, creating modes of signification. This process demonstrates the machine's capacity for nuanced interpretation of cultural and aesthetic codes, effectively bridging the gap between textual input and musical output within the broader cultural semiosphere.

Beyond mere generative capabilities through prompts, Udio's functionality extends to the transformation of existing audio based on textual descriptions.[8] This feature enables a wide range of audio metamorphoses: the system can transpose an input across diverse genres, augment instrumental tracks with vocal elements, enhance rhythmic patterns with melodic structures, and facilitate numerous other sonic transformations. For instance, Udio can convert a minimalist electronic beat into a full-fledged orchestral piece, or infuse an acoustic ballad with elements of noise, all guided by descriptive text. Such versatility allows Udio to render both newly generated and transformed audio within contextually relevant musical signifiers, as delineated by the textual prompt. This advanced functionality showcases Udio's proficiency in not only synthesizing music that aligns with user-specified parameters but also in reinterpreting existing audio to encapsulate the stylistic and instrumental essence conveyed in the descriptive text.

**2.2 Semiotic Transduction in Text-to-Audio: A Multi-Stage Process Analysis**

The text-to-audio modality represents a sophisticated form of semiotic transduction[9], bridging linguistic input and audio output through a series of transformative stages. This process not only demonstrates technological innovation but also illuminates key concepts in semiotics, cognitive science, and information theory. The following analysis briefly deconstructs the operational framework of text-to-audio, elucidating each stage through both technical explanation and theoretical contextualization.

**2.2.1 Operational Dynamics of Model Training: From Raw Data to Semantic Representation**

**Raw Audio Data → Feature Extraction**

The system starts with a large dataset of raw audio. It analyzes this data to extract key features such as frequency spectra, rhythmic patterns, and timbral characteristics. This step reduces complex waveforms to a set of manageable, meaningful attributes such as tempo. In this initial stage, the vast complexity of raw audio waveforms is reduced to a set of salient features. This process, akin to what Goodman (1968) terms "exemplification" in his theory of symbols, involves a selective abstraction of relevant audio characteristics. This stage resonates with Gibson's (1979) concept of affordances, where the environment (in this case, the audio data) is perceived in terms of its action possibilities.

**Feature Representations → Latent Space Encoding**

The extracted features undergo a crucial transformation as they are encoded into a high-dimensional latent space. This latent space serves as a compact, continuous representation where audio characteristics with semantic or acoustic similarities are proximally located. The encoding process in modern text-to-audio models typically employs transformer architectures working on audio tokens. These models, such as Udio, likely utilize a system similar to the CLAP (Contrastive Language-Audio Pretraining) model for handling tags and metadata, enabling a sophisticated bridging of linguistic and audio domains.

The latent space becomes a complex interpretant of the audio features, losing direct correspondence to audible properties but gaining in associative and generative potential. This abstraction aligns with Derrida's (1967) concept of "différance," where meaning emerges from the relationships and differences within a system rather than from direct representation.

**Latent Space → Text Association**

Through training on audio and text data, the system then learns to associate regions in the latent space with textual descriptors through training on paired audio-text data. These associations encompass both genre labels (e.g., "jazz", "ambient") and more nuanced semantic descriptors (e.g., "melancholic", "ethereal"). This process exemplifies intersemiotic translation, where signs from one semiotic system (text) are interpreted through another (abstract audio representations).

The formation of these associations involves what Fauconnier and Turner (2002) describe as conceptual blending, where elements from disparate domains (language and abstract audio representations) are integrated to form new conceptual structures. This blending process is crucial for the system's ability to generate or transform audio based on textual inputs, as it creates a bridge between linguistic concepts and sonic characteristics within the latent space.

**2.2.2 Operational Dynamics from User Input to Audio Generation**

**Text Input → Latent Space Navigation**

When a user provides a text input, the model employs it to navigate the latent space, identifying regions that best match the textual description and considering both genre tags and semantic descriptors. It is in this stage that the model's behavior becomes particularly intriguing. The system does not simply locate a single point in the latent space but rather identifies regions or clusters that correspond to the input descriptors.

Notably, the model's navigation exhibits a degree of variability and nuance:

a) Weighted Influence: Certain words in the input may exert a stronger influence, causing the model to gravitate more heavily towards specific clusters associated with those terms. For instance, in the input "melancholic ambient with subtle jazz influences", the term "ambient" might dominate the latent space selection, with "melancholic" and "jazz" having secondary, modulating effects. When given a textual prompt, the model doesn't simply match it to a single point in the latent space but instead activates a cluster of related concepts. For instance, a prompt including "ethereal" and "nocturnal" might activate a region of the latent space associated with Ambient Pop, but also draw influences from areas linked to "mysterious" and "atmospheric" qualities.[10]

b) Interpolation and Synthesis: When presented with a combination of descriptors that map to disparate regions of the latent space, the model attempts to synthesize a recombined position. This interpolation between distinct clusters can lead to highly idiosyncratic results, effectively creating an interpolated "region" that blends characteristics of multiple sonic identities.[11]

When a user provides a textual input, it initiates a navigation through the latent space. This process can be viewed through the lens of Eco's (1984) "encyclopedia" model of semiotics, where the AI system's learned associations form a vast network of potential associations. The navigation might lose some of the nuanced linguistic meaning of the input but gains in its potential to generate novel audio outputs. This stage also resonates with Deleuze and Guattari's (1980) concept of the "rhizome," where any point can be connected to any other, allowing for non-hierarchical and multiple entryways into the generation process. Much like a rhizome, the model's data points and influences are not arranged linearly but instead form a network of multiple entry and exit points. This allows for transmutational creativity, as the model accesses and synthesizes disparate musical influences — from classical piano to drumming to modern electronic synths — without a pre-determined pathway or a dominant root. Evidence of this tremendous creative ability is the model's ability to not repeat the same generation of its textual prompt, ensuring each output is distinct and reflective of the diverse potentialities within its dataset.

This multi-layered indeterminacy creates what we might call a "cascade of aleatory events," each influencing the next in a complex chain of interpretation and realization. The result is a form of musical composition that is simultaneously deterministic (in that it follows the learned patterns and associations of the model) and deeply indeterminate (in that the specific realization is unpredictable and unrepeatable). This is evident in the absence of repetitive outputs for identical prompts. For instance, if we input " ambient pop, glitch, folk, classical, no vocals " multiple times, each generated piece is distinct, showcasing the model's expansive creative capacity and its ability to explore different combinations within its latent space.[12]

**Latent Space Selection → Audio Generation**

Once the appropriate regions of the latent space are "selected", the model uses this information to generate new audio. It translates the latent space coordinates back into audio features and then into a full audio waveform. The model translates latent space coordinates into a full audio waveform using advanced synthesis techniques. This process goes beyond simple sound generation, incorporating nuanced audio characteristics that professional music production typically requires. As part of the generation process, the model implicitly performs tasks akin to mixing and mastering. It balances various audio elements, adjusts equalization, and applies appropriate dynamic processing, often resulting in well-produced sound.

This process exemplifies Simondon's (1958/2017) concept of "concretization," where abstract potentials are actualized into specific, material forms. The transformation from latent space to audio waveform represents a critical juncture in the semiotic transduction process, where the system bridges the gap between abstract representations and perceivable sonic realities. Moreover, this stage can be understood through the lens of Peirce's (1931-1958) theory of semiosis, particularly his concept of the "interpretant." Text-to-audio models act as both sign-makers and sign-interpreters, translating textual signs into musical signs. The generated audio serves as a complex interpretant of both the original textual input and the latent space representation, creating a new sign that can itself be subject to further interpretation and analysis.

The potentiality and actuality of this latent space are reminiscent of Deleuze's (1968) concepts of the virtual and the actual, where the virtual (latent space) contains a multiplicity of potentials, and the actual (generated audio) represents a specific realization of these potentials. The audio generation stage thus represents a moment of creative synthesis, where the abstract possibilities encoded in the latent space are transformed into tangible sonic experiences, bridging the gap between linguistic description, mathematical representation, and auditory perception.[13] Notably, the model often demonstrates a remarkable ability to capture the intangible qualities or "vibe" suggested by the text prompt.[14] while at times also producing outputs that significantly diverge from the intended aesthetic or semantic context—resulting in outputs that may seem incongruous or misaligned with the user's expectations.[15]

**2.3 Prompt Interpretation and Output Diversity in Text-to-Audio Synthesis**

The core of contemporary text-to-audio models' capability lies in their advanced natural language processing (NLP) algorithms. These algorithms are not merely parsing text for direct translation into sound but are equipped with a rich understanding of language semantics. They interpret adjectives, metaphors, and contextual cues, going beyond literal descriptions to grasp the intended emotive or atmospheric nuances that the user wishes to convey. This semantic understanding allows the models to match auditory outputs with the descriptive text's mood, genre, or atmosphere accurately[16] – often exhibiting a gestalt fluency.

These models operate within a framework of "semantics in disguise," interpreting textual inputs through complex networks of metatags and associations. This process is exemplified in how these models handle varying degrees of prompt complexity. For instance, when presented with a straightforward description such as "a tranquil morning in a serene forest,"[17] the model discerns the peaceful qualities suggested by "tranquil" and "serene," translating them into an audio output characterized by harps, flutes, and flowing melodies. Models like Udio further refine this process by generating their own intermediary tags—in this case, potentially "calm," "mellow," "new age," and "ambient"—offering insight into their semantic categorization mechanisms.

The nature of the prompt plays a crucial role in determining the character of the output. When prompts align with common associations, the model tends to produce generalized audio outputs that meet broad expectations. However, the true versatility and potential for innovation in these systems become apparent when users provide more particular or transgressive prompts. For example, a complex prompt like "john oswald plunderphonics experimental, deconstructed, classical music, stockhausen, no vocals" prompts the system to navigate a more intricate semantic landscape. Such a prompt leads to the association of metatags like "Instrumental, Avant-garde, Quirky, Eclectic, Experimental, Playful, Electronic, Humorous, Political, Concept album, Spoken word, Plunderphonics"[18] or "no vocals, classical, deconstructed, 2010s, plunderphonics, experimental, electronic, abstract, eclectic, melodic, alternative r&b, instrumental."[19] This complex interplay of semantic cues drives the model to venture into different clusters of its latent space, resulting in outputs that navigate unconventional musical structures and expectations. These models do not simply translate text to sound; they engage in peculiar interpretations of semantic cues, transforming them into particular sound objects.

The model's ability to decode and recontextualize semantic information within the audio domain facilitates a potential for destabilization and recontextualization of musical concepts. By incorporating terms like "deconstructed" or "plunderphonic," users can guide the model towards more experimental territories, pushing it to produce distinctive outputs that diverge from typical musical norms. Notably, certain words act as potent semantic triggers, propelling the model into disparate and often highly creative directions. This phenomenon reveals a complex landscape of semantic interplays, where specific linguistic cues can lead to particular aesthetic outcomes—a rich territory that invites further exploration and mapping. This capability underscores the model's potential not just as a tool for replicating familiar sounds, but as a medium for exploring boundaries of musical expression, offering a unique lens through which to examine the relationship between language, conceptualization, and musical creativity.

Udio's ability to navigate and synthesize within its latent space is particularly noteworthy. It allows for the generation of audio that not only corresponds to explicit textual descriptors but also captures subtle interactions between these descriptors. This can result in outputs ranging from highly archetypal (closely matching established clusters) to remarkably idiosyncratic (requiring significant interpolation between disparate regions). Udio, in particular, showcases a remarkable potential to traverse and navigate sonic clusters and signs through its synthesis methods. This capacity for nuanced interpretation and creative synthesis often produces results that challenge simple categorization, illustrating the potential of text-to-audio as a medium for exploring sonic territories at the intersection of established auditory concepts.

**2.4 Cognitive Assimilation and Intersemiotic Musicking**

Drawing from Jakobson's (1959) concept of intersemiotic translation and Small's (1998) notion of musicking, intersemiotic musicking refers to a particular form of musical engagement that entails the interpretation and transformation of signs across different semiotic systems in the context of music creation and perception. This encompasses not only the translation of textual signs into sonic outputs, as seen in text-to-audio systems, but also other forms of intersemiotic translations in music, such as timbre transfer, visual-to-audio mappings, or gestural-to-sound translations. It refers to the act of musicking through various modes of cross-modal musical interaction and creation, facilitated by AI and other technologies. This process creates unique semiotic spaces where diverse modes of expression converge, offering distinct pathways for musical creativity and perception.

Within the text-to-audio modality, a specific form of intersemiotic musicking occurs as users engage in translating musical ideas into textual prompts. This process is not merely a technical operation but a complex semiotic interaction that involves both the user and the AI system. The subsequent interaction between AI outputs and listener expectations demonstrates a form of semantic and semiotic assimilation. Listeners, drawing from their knowledge and expectations, often imbue AI-generated outputs with attributes they associate with the prompt, even when these attributes are not explicitly present in the output. This phenomenon aligns with Eco's (1976) theory of semiotics, where signs function within a network of cultural codes and subcodes. In the case of text-to-audio, users must navigate not only their own understanding of musical concepts but also anticipate how the AI system might interpret these linguistic signs.

For instance, when presented with a MusicLM interpretation of "saxophone glenn gould plays bach counterpoint"[20] listeners might perceive elements of Gould's style, even if the AI's output lacks specific nuances of his playing. This cognitive process demonstrates a form of 'associative projection,' where the listener actively applies associations to the AI's output based on the prompt's keywords, even when these associations aren't inherently present in the sound.

Upon review of the model's output, one discerns the presence of Bach's compositional structures and the distinctive sonority of the saxophone. However, the nuances characteristic of Glenn Gould's playing — his unique touch and interpretive timing — are notably absent. This absence is not due to a deficiency in the model but stems from its lack of exposure to Gould's performances in its dataset. Yet, in the cognitive interplay of expectation and perception, an interesting phenomenon occurs. The listener, aware of Gould's stylistic legacy, might perceive the AI's interpretation as something that could be aligned with Gould's ethos, thus experiencing a form of semantic and semiotic assimilation. Listeners, drawing from their knowledge and expectations, may imbue the output with attributes they associate with the prompt, even if those attributes are not explicitly present in the AI's output. This phenomenon of prompt-guided constructive listening illustrates how the initial textual signs guide the listener's interpretation, leading to a particular, albeit potentially misattributed, listening experience.

**3. Cognitive Frameworks in AI-Mediated Musicking: Structural Paradigms, Schema Dynamics, and Post-Structural Perspectives**

**3.1 Structuralism: Music, Meaning, and Cultural Constructs**

Structuralism provides a lens through which we can analyze how text-to-audio models both conform to and challenge existing musical paradigms.

Structuralism, a salient intellectual movement of the mid-20th century, contends that the myriad aspects of human culture must be perceived within a systemic framework of signs and relational structures, as postulated by Ferdinand de Saussure (De Saussure, 1916). In the realm of music, this paradigm shifts the focus from viewing musical works as standalone entities to interpreting them as integral parts of an intricate web of significations and operational roles within a broader cultural matrix. Structuralist thought posits that human culture is understood through structures—predetermined patterns and relationships that define our perceptions and realities. In the context of music, structuralist perspectives lead listeners to seek out patterns and forms that align with established musical structures. These structures could be harmonic progressions, rhythmic patterns, genre-specific conventions and aesthetic signifiers.

These foundational components coalesce to imbue music with its profound capacity to convey nuanced meanings and reflect the cultural milieu from which it emerges. Our contemporary practices of musicking are predicated upon and informed by these structuralist frameworks, which delineate the systematic intricacies of music's language and fundamentally shape our engagement with music, whether through creation, performance, or listening.

Within the domain of music, structuralism posits a kind of territorialization of auditory constructs, suggesting that musical components are organized within a circumscribed sphere of rules, conventions, and ingrained patterns. This structuralist perspective leads to a cognitive territorialization where the mental and perceptual realms are cultivated by, and enmeshed with, established musical narratives and languages. Our auditory sensibilities are so ingrained with these traditional constructs that sounds outside these parameters often go unrecognized or unappreciated. Consequently, the structuralist framework engenders meta-narratives (Lyotard, 1979) in music—those predominant themes and theoretical structures that provide coherence yet potentially constrain or sideline non-conforming musical expressions. Within this view, musical syntax, theoretical frameworks, instrumental connotations, and genres serve as meta-narratives, fortifying our musical experiences with coherence, legitimacy, and validity.

Moreover, our semantic predicaments—bound by the limitations of our linguistic frameworks—shape our reception of AI-generated outputs. As AI interprets textual prompts through a schematic lens to produce music according to its data and design, listeners similarly apply their schematic understanding to ascribe meaning to the music they hear according to their structuralist frameworks. These predicaments manifest when audiences engage with music that has originated from textual prompts, anticipating that the music will mirror the semantic content and emotive underpinnings of the language that informed its creation. The extent to which AI-synthesized compositions align with or diverge from these semantic expectations significantly impacts the listeners' receptivity and valuation of the musical work.

In the light of Deleuzian philosophy, particularly as articulated in 'A Thousand Plateaus', such musical encounters evoke an arborescent mode of thinking, where every piece of music is expected to trace back to an identifiable origin or a semiotic 'root' (Deleuze & Guattari, 1980). This arborescent expectation underscores our tendency to seek familiarity and coherence, to trace music back to something preconceived and structured within our mental constructs of music.

**3.2 Schema Dynamics in AI-Mediated Musicking: Assimilation and Accommodation**

In the context of music perception, listeners approach musical experiences with pre-existing schemas—mental structures shaped by lifelong exposure to musical patterns, cultural contexts, and emotional associations. These schemas form a complex web of expectations about what music is and how it should sound. A schema is a cognitive structure that represents a collection of ideas and ways of organizing knowledge. It's how an individual understands the world and organizes information—it's essentially a mental blueprint or set of rules that are used to interact with and make sense of the world. Every listener approaches music with a set of cognitive and cultural schemas—mental structures that organize past experiences and provide a framework for understanding future events. These schemas are the result of lifelong exposures to musical patterns, cultural contexts, and emotional associations, forming a complex web of expectations and understandings about what music is and what it should sound like.

Piaget's concepts of schema assimilation and accommodation (Piaget, 1957) provide a valuable framework for understanding the cognitive processes involved in perceiving AI-generated music. Listeners approach musical experiences with pre-existing schemas—mental structures shaped by lifelong exposure to musical patterns, cultural contexts, and emotional associations.

In the context of text-to-audio, both listeners and AI models engage in forms of schema-based processes:

**1. Assimilation:** In Piaget's theory of cognitive development, schema assimilation is the process by which a person incorporates new information into an existing mental framework, or schema (Piaget, 1957). Users approach these tools with preconceived notions of what music 'should' sound like, which are shaped by their cultural background, education, and personal experiences. As users input prompts into the AI system, they often draw upon familiar musical concepts and signifiers, expecting the AI to produce outputs that align with their established schemas. Text-to-audio models like Udio perform similar mappings, contextualizing text prompts within pre-learned schemas to generate music. For example, a prompt like "jazz in a smoky bar" activates schemas related to jazz music and intimate atmospheres. meeting the listeners' anticipations by aligning the generated music with familiar auditory patterns. This interaction typifies the assimilation process: the musical outputs are perceived through and incorporated into the listener's existing cognitive and cultural schemas.

**2. Partial Schema Assimilation:** Udio's ability to blend genres and create distinct recombinations often results in "partial schema assimilation," where some elements fit existing categories while others remain unassimilated. This echoes Huron's (2006) ITPRA theory of expectation, where listeners' predictions are partially met and partially violated. The textual prompts that generate the music add another layer to this assimilation process. Listeners must negotiate between their expectations based on the linguistic input and the actual sonic output, creating what Leman (2008) might describe as a complex "action-perception coupling" between language, expectation, and auditory experience.

**3. Accommodation:** Accommodation, contrastingly, is invoked when existing schemas are inadequate to process new information. It involves modifying or constructing new schemas to assimilate this information. This cognitive adjustment is more profound and transformative, resulting from a significant dissonance between existing schemas and new stimuli. In musical terms, accommodation might occur when a listener encounters a piece that defies genre conventions or introduces odd combinations of structural elements, thereby challenging the listener's preconceived notions of musicality. This process aligns with what Pressing (1988) describes in his theory of improvisation as the development of new "event clusters" or cognitive representations of musical patterns.

The presence of textual prompts in this process adds another layer of cognitive complexity. Listeners engage in what Koelsch (2011) describes as the "simultaneous integration of syntax in music and language," but with the added dimension of the AI interpretation. This creates a triadic relationship between linguistic input, AI processing, and listener cognition, each influencing and being influenced by the others in a complex feedback loop. Moreover, the constant negotiation between textual prompts, AI interpretation, and sonic output may foster what Bourriaud (2002) terms "relational aesthetics,"[21] where meaning and artistic value emerge from the complex interactions between human language, machine learning, and human perception.

When engaging with music, we are not passive recipients; we actively endow the music with significance, filtered through the lens of our accrued knowledge and experiential backdrop. This dynamic is particularly intensified in the context of text-to-audio interactions where the conventional boundaries and distinctions may not just be stretched but subverted or bypassed entirely. Often, music that is deeply rooted in recognized forms and conventions may be perceived as 'correct', while departures from these norms may be dismissed as erroneous or cacophonous.  The profound contribution of text-to-audio models such as Udio is most salient when their outputs engage with and transfigure the listener's preconceived cognitive structures. Such encounters often induce a creative dissonance, a fruitful discord that necessitates a reevaluation or acknowledgement of entrenched auditory expectations. This cognitive recalibration prompts an expansion of the listener's mental schemas to accommodate these novel musical forms. As listeners realign their auditory sensibilities to integrate these unexpected sounds, there is potential for an awareness of musical appreciation and the cultivation of emergent aesthetic sensibilities.

The paralleled cognitive processes of assimilation and accommodation, as delineated by Piaget, are not only instrumental for understanding human development but also offer a compelling framework for analyzing the interactivity between users and text-to-audio models. This dynamic is bidirectional; both the users and the AI systems undergo analogous processes, albeit in different contexts and by virtue of different mechanisms.  The conceptual blending of humans assimilates musical concepts with linguistic expressions, often drawing on emotional and cultural associations, while AI models perform a more systematic mapping between textual and musical features based on learned patterns from tags. When these concepts are applied to the interaction with text-to-audio AI systems, it becomes evident that users mostly engage with these technologies through the lens of structuralist thought and schema assimilation. They expect the AI to produce music that conforms to certain rules and structures they understand. However, as post-structuralist thought would suggest, the sound objects created by these models can often defy these expectations, offering outputs that are not easily categorized within the traditional musical structures - prompting a re-evaluation of these pre-existing schemas, fostering an inferentialist intuition in musical perception.

Post-structuralist thought, with its emphasis on the instability of meaning and the multiplicity of interpretations, provides a complementary framework for engaging with text-to-audio AI in a creative and critical manner. It encourages us to move beyond mere reflection on established musical structures and our own listening habits, pushing us to actively deconstruct and reimagine the fundamental concepts of music, creativity, and meaning-making. This approach is particularly pertinent in the context of AI-generated music, where traditional notions of authorship, genre, and musical coherence are already being challenged.

### 3.3 Post-Structuralism; Ambiguity, Multiplicity and Manifold Reinterpretation

Post-structuralism, which emerged as a response to structuralism, argues that meaning is not fixed and is subject to the interplay of various contexts and perspectives. It emphasizes the inherent instability in language and the incapacity of any single structure to encapsulate the nuances of human experience fully (Réda, 2005). In stark contrast to structuralism's focus on fixed relationships and defined meanings within a system, post-structuralism embraces ambiguity, multiplicity, and the potential for manifold reinterpretation.

Challenging the self-containment of structural systems, post-structuralism, in affinity with postmodernism, disputes the adequacy of any fixed set of structures to encompass the breadth of human and artistic experience, including the variegated landscape of musical expression. It promotes the departure from rigid, hermetic structures towards more permeable, interpretative frameworks. In doing so, it aligns with a postmodernist ethos that values indeterminacy, diversity, and reinterpretation over fixed relationships and definite meanings within a system.

Within the milieu of music, post-structuralism actively deconstructs established musical hierarchies and categorizations, forging a path for the investigation and valorization of musical forms previously marginalized or undervalued by structuralist criteria. As AI continues to infuse the musical domain, it compels a reevaluation of existing structural and semantic schemas. This evolution calls for an engagement with music that is explorative and, at times, adversarial—challenging stable identities and signifiers and embracing variance and aesthetic drifts. An adversarial approach to 'musicking' is essential to truly engage with and appreciate the richness of the musical interstices unveiled by these models. This willingness to confront and contest the traditional paradigms enables a profound exploration of the liminal spaces within music and the latent spaces in these models, wherein ambiguity, manifold reinterpretation and plurality thrive.

Many of Udio's outputs such as "john oswald plunderphonics experimental, deconstructed, classical music, stockhausen, no vocals" demonstrated above, when analyzed through a post-structuralist lens, often demonstrate a potential for divergent creation that challenges established representational norms in music. By embracing AI models' capacity to deviate from conventional forms, we open avenues for novel aesthetic trajectories and sonic expressions. This approach moves beyond the limitations of arborescent thinking, instead fostering a rhizomatic multiplicity that expands the boundaries of musical signification. Text-to-audio models like Udio are particularly compelling when their outputs create a productive tension with listeners' preconceived structures, inducing a creative dissonance that challenges auditory expectations. This cognitive recalibration, coupled with a post-structural attitude, can lead to an expansion of mental schemas and the cultivation of distinctive aesthetic sensibilities, ultimately enriching understandings and appreciations of music in the age of AI.

### 3.5 Metacognition and Reflective Listening

Flavell's (1979) concept of metacognition – thinking about one's own thought processes – is particularly relevant to the experience of this modality. Listeners may not only perceive the music but also actively reflect on the AI's interpretative processes and their own responses to these interpretations. This metacognitive engagement can be further refined into what we might call "structural-aware listening." This form of listening involves not only an awareness of the AI's interpretative processes but also a conscious reflection on the listener's own structural predispositions and how they shape the perception and interactions with these models. Drawing on Bourdieu's (1977) concept of habitus, we can understand these predispositions as internalized structures, unconsciously acquired through sociocultural experiences, that guide our perception and interpretation of music.

In the context of text-to-audio, structural-aware listening might encompass:

**1.** Recognition of familiar musical structures and how they've been reinterpreted or subverted by the AI.

**2.** Awareness of one's own expectations based on the textual prompt and how they align or conflict with the AI's output.

**3.** Reflection on the cultural and historical contexts that shape one's interpretation of both the prompt and the resulting music.

**4.** Evaluation of how personal musical training (or lack thereof) impacts perception of AI compositional choices

This form of metacognitive engagement aligns with forms of reflective and critical listening, but extends it to include awareness of the technological mediation inherent in AI-generated music. It creates a multi-layered listening experience where the listener simultaneously engages with the music, reflects on their own cognitive processes, and considers the AI's role in the creative process. Furthermore, by adopting a post-structuralist perspective, we can enrich and challenge this reflective practice. While metacognition enables us to examine our thought processes and predispositions, post-structuralism urges us to question the very structures and meanings we take for granted in music. This combination of metacognitive awareness and post-structural critique creates a powerful framework for engaging with these interactions, potentially leading to new understandings of musical creativity, perception, and the evolving relationship between human listeners and text-to-audio outputs.

**4. Text-to-Audio models as Epistemic Quasi-Objects**

The text-to-audio modality presents a unique paradigm that challenges and expands established notions of musical cognition, cultural codification, and approaches to creating and perceiving music. These AI models function as what Serres (1982) terms "quasi-objects" – entities that mediate between subject and object, human and non-human. As quasi-objects, they occupy a dynamic intermediary position, constantly shifting between roles as they facilitate the complex process of translating text into sound.

This technology, by intertwining linguistic expression with sonic generation, opens up idiosyncratic potentials for musical exploration and creation. The resultant audio outputs are products of a nuanced negotiation, reflecting the quasi-object nature of the model – neither purely human nor entirely machine, but a unique hybrid that transcends traditional categorizations. These models function as powerful epistemic tools, revealing embedded assumptions about music and sound. As Ihde's (1979) "hermeneutic technics," they do more than translate between text and sound; they serve as active interpreters, making explicit the implicit structures that underlie our musical understanding.

Text-to-audio models straddle a complex dichotomy. These tools can act as both magnifying glass and scalpel, enlarging our view of semiotic processes while dissecting the composite layers of meaning within musical traditions. On the one hand, they risk perpetuating the dominant cultural signifiers embedded within their training data, yet on the other hand, they offer a medium for deconstructing (Derrida, 1967), and repurposing these very signifiers. Through these tools, we can gain a profound awareness of the cognitive processes underlying our interaction with music and language, offering a profound awareness of the inferentialist processes that shape our musical engagement.

As we engage with these technologies, we are called to participate in the construction of new interactions, and psychological and representational maps, fostering the development of previously inconceivable musical aesthetics and expressions. To unveil new horizons within the musical domain, we must move beyond representational and arborescent models that bind us to established trajectories. Instead, we should embrace the aesthetic drifts and infra-significances that AI can offer, much like how the drum machine, once criticized for its imperfect emulation of drum sounds, became the progenitor of entirely new musical forms.

In conclusion, text-to-audio AI models, as epistemic quasi-objects, offer a unique lens through which we can examine and reimagine musical interaction and research. They serve not only as tools for creation but as mirrors reflecting our cognitive processes, cultural assumptions, and the very nature of musical meaning-making. By engaging with these models critically and creatively, we open new avenues for understanding and expanding the boundaries of musical expression in the age of AI.

**5. Conclusion**

In discussing the parallels and divergences between human cognitive processes and AI's interpretative mechanisms, this paper highlights the confluence of cognition, structuralism, and inference in music perception and creation. The interface with text-to-audio models is not just a tool for generating sound but a means of probing the depths of how we discern meaning, emotion, and organization in music, influenced by our cultural and linguistic contexts. By exploring the similarities and divergences between AI-generated music and human musical intuition, this paper contributes to dialogues on aesthetics, cognition, artificial intelligence, and the arts. It invites us to rethink the demarcations of musical meaning attributed to humans and machines - highlighting the ability of text-to-audio to transgress and navigate established musical forms.

As epistemic tools, these models can act as catalysts for a deeper understanding of the semiotic processes in music, serving as reflective instruments that bring the interpretative interplay of language and sound to the forefront. As meta-objects, they compel us to reevaluate established musical structures and to reconsider the instrumentality or instrufacility of AI in music creation. As state-of-the-art models develop, they serve as structural reflectors – bringing to the fore understandings of language and signifiers in music, offering new perspectives on semiotic and semantic mechanisms.

As we interact with these tools, we not only encounter a new realm of musical possibilities but also come face to face with a critical mirror —one that reflects our position within an intricate cultural schema and incites us to either accommodate or reconfigure it. The text-to-audio paradigm emerges, therefore, not merely as a tool but as an exceptional instrument of epistemic and aesthetic inquiry in the grand schema of music - modulating our understanding of music's structure and its semiotic nuances, marking a significant epoch in the evolution of musical interactions.

## Footnotes

1. Youtube Dreamtrack Demonstration: https://www.youtube.com/watch?v=1uW_AUwEv-0 ↩
2. King Willonius - BBL Drizzy: https://www.youtube.com/watch?v=1uW_AUwEv-0 ↩
3. Houghton, B. (2023, July 26). Metro Boomin Clarifies That Viral 'Future' Track Is AI-Generated. Billboard. https://www.billboard.com/business/tech/metro-boomin-bbl-drizzy-future-ai-sampling-1235682587/ ↩
4. Udio Output of "extended techniques, free jazz, avant-garde, experimental": https://drive.google.com/file/d/1I3zLx5RbMupub9stfW4VFBzouGMuyDJR/view?usp=drive_link ↩

5. Udio Output of "A song about a car crashing into a musical instrument shop, geek rock, hard rock":
https://drive.google.com/file/d/1ebClvgdftD_xXyw9jhFFka64FCtY2lW6/view?usp=drive_link ↩

6. Udio Output of "Phillip Glass, Einstein on the Beach" :
https://drive.google.com/file/d/1vXrw5ZxO61QECYCGjvSfgIGCMm4v-7Zj/view?usp=drive_link ↩

7. Udio Output of "Cinematic, emotional, orchestral, D Minor, 100 BPM, no vocals":
https://drive.google.com/file/d/1NxpNYlR-vCdJkxEouWck3WK5fBZHg6A7/view?usp=drive_link ↩

8. Udio Audio Remix feature Demonstration: https://twitter.com/udiomusic/status/1818719659157602469 ↩

9. Transduction, a term introduced by Gunther Kress (1997) in his social semiotic approach to multimodality, refers to the process of transforming meaning from one mode of expression to another. ↩

10. Udio Output of "melancholic ambient with subtle jazz influences":
https://drive.google.com/file/d/1anJ6l25gJHayOyFxfD0ZZBNBwDIXhnPt/view?usp=drive_link ↩

11. Udio Ouput of "A deconstructed baroque harpsichord solo with dubstep elements":
https://drive.google.com/file/d/1Fc5ICWySi0JDmRS1BFs-_yX0CXY-TrhW/view?usp=drive_link ↩

12. Udio Outputs of "ambient pop, glitch, folk, classical, no vocals":
https://drive.google.com/drive/folders/1U0rTKG1F6lBqUm-3k-u7KDQRkRrSUjyj?usp=drive_link ↩

13. Different text-to-audio models employ varied training approaches. MusicLM, using the MuLan model, is trained on a dataset without explicit text-audio pairings, enabling it to respond to complex prompts through cross-domain generalization (Agostinelli et al., 2023). In contrast, models like Udio and Suno reportedly use extensive music collections from platforms like YouTube, along with associated tagging systems and descriptors, similar to those found on websites like RateYourMusic. This approach leverages diverse online musical content and user-generated descriptors, resulting in more versatile outputs with higher sonic fidelity. The quality and diversity of training data are crucial for robust performance across various musical styles and genres. However, the cultural and linguistic embeddedness of these systems raises important questions about appropriation, representation, bias, and the nature of musical universals in AI-generated creativity. ↩

14. Udio Output of "R&B, Trey Songz, sexy song wanting to be with a woman":
https://drive.google.com/file/d/1dJV-cdTK5mM9JHZAScWEB8A0uOwhwh-_/view?usp=drive_link ↩

15. 3. Udio Output of "Instrumental, Avant-garde, Quirky, Eclectic, Experimental, Playful, Electronic, Humorous, Political, Concept album, Spoken word, Plunderphonics":
https://drive.google.com/file/d/1J7vBaFB2rDPWk7XbIQuBP9p-veB3eA2c/view?usp=drive_link ↩

16. The term "accurately" here does not imply an objective correctness but rather a high degree of fidelity to the fluency at play, revealing the model's interpretive acumen of its latent space. ↩

17. Udio Output of "a tranquil morning in a serene forest tags – calm, mellow, new age, ambient which makes sense": https://drive.google.com/file/d/1abMdUAEPurPVM02vDygxxq7h10dSfkhn/view?usp=drive_link ↩

18. Udio output of "john oswald plunderphonics experimental, deconstructed, classical music, stockhausen, no vocals": https://drive.google.com/file/d/1MVp75EtyWlFAEvpJOl-xtkupiY_UZR_p/view?usp=drive_link ↩

19. Udio output of "john oswald plunderphonics experimental, deconstructed, classical music, stockhausen, no vocals": https://drive.google.com/file/d/1vmgGIjI05QW2lQq7a4VDLQgRTi_ghIM9/view?usp=drive_link ↩

20. MusicLM output of "saxophone glenn gould plays bach counterpoint": https://drive.google.com/file/d/1kZlHpGBGaPdWr2C87UkH6So_FNLlzrsC/view?usp=drive_link ↩

21. Defined by Bourriad as "a set of artistic practices which take as their theoretical and practical point of departure the whole of human relations and their social context, rather than an independent and private space." (Bourriard, 2002) ↩